\documentclass{article}

\usepackage{arxiv}

\usepackage[utf8]{inputenc} 
\usepackage[T1]{fontenc}    
\usepackage{hyperref}       
\usepackage{url}            
\usepackage{booktabs}       
\usepackage{amsfonts}       
\usepackage{nicefrac}       
\usepackage{microtype}      
\usepackage{lipsum}
\usepackage{graphicx}
\usepackage{makecell}
\usepackage{amsmath}
\usepackage{mathtools}

\usepackage[colorinlistoftodos]{todonotes}

\usepackage{color}

\title{Acceleration of Radiation Transport Solves Using Artificial Neural Networks}

\author{
  Mauricio Tano \\
  Department of Nuclear Engineering\\
  Texas A\&M University\\
  College Station, TX 77840 \\
  \texttt{mtano@tamu.edu} \\
   \And
  Jean Ragusa \\
  Department of Nuclear Engineering\\
  Texas A\&M University\\
  College Station, TX 77840 \\
  \texttt{jean.ragusa@tamu.edu} \\
}

\begin{document}
\maketitle

\begin{abstract}
Discontinuous Finite Element Methods (DFEM) have been widely used for solving $S_n$ radiation transport problems in participative and non-participative media. In the DFEM $S_n$  methodology, the transport equation is discretized into a set of algebraic equations that have to be solved for each spatial cell and angular direction, strictly preserving the following of radiation in the system. At the core of a DFEM solver a small matrix-vector system (of 8 independent equations for tri-linear DFEM in 3D hexehdral cells) has to be assembled and solved for each cell, angle, energy group, and time step. These systems are generally solved by direct Gaussian Elimination. The computational cost of the Gaussian Elimination,  repeated for each phase-space cell, amounts to a large fraction to the total compute time. Here, we have designed a Machine Learning algorithm based in a shallow Artificial Neural Networks (ANNs) to replace that Gaussian Elimination step, enabling a sizeable speed up in the solution process. The key idea is to train an ANN with a large set of solutions of random one-cell transport problems and then to use the trained ANN to replace Gaussian Elimination large scale transport solvers. It has been observed that ANNs decrease the solution times by at least a factor of 4, while introducing mean absolute errors between 1-3 \% in large scale transport solutions.
\end{abstract}


\section{Introduction}

Radiative transfer deals with the transport of particles in a background medium, which absorbs, emits and scatters these particles \cite{chandrasekhar2013radiative}. This process is of great transverse importance across science and engineering. For instance, it describes the heat transfers taking place in semitransparent stellar and planetary atmospheres. Moreover, it describes the transport of neutrons and photons in nuclear reactors, as well as in shielding and medical diagnose applications. The radiative transfer equation (RTE) is a complicate linear integral-differential equation, which can be solved analytically only for very simple cases. Hence, there have been historically a large effort to develop numerical methods that can solve this equations across different media, which lead to a large set of numerical methods used for its solution. 

Among these methods, the Discontinuous Galerkin(DG) Finite Element Method (FEM) was first introduced by Reed and Hill \cite{reed1973triangularmesh} and Lesaint and Raviart \cite{lesaint1974finite} in the early 1970s. This is a local Finite Element (FE) method, where the inner-element continuity in the discretized grid is released and the shape functions are constructed on each element. Compared to others grid-based schemes, DGFEM combines the salient features of both the continuous finite elements and the finite volume methods. Furthermore, this method is generally combined with a discretization in discrete ordinates in the angular direction, a binned discretization in the energy dependence, and a discretization in time for transient problems \cite{adams2001discontinuous}. By examining the space dependence, DGFEM reduces the RTE problem to a balance problem in one computational cell, which is connected to its neighbours by the radiative fluxes at interfaces. This problem is generally solved by direct Gaussian elimination given the small number of equations involved. However, it needs to be solved for all cells in the domain, for each of the discrete directions and energies and, eventually, for each time step in transient problems. The solution of these systems is the core of the computational cost of DGFEM for radiation transport . Nevertheless, the construction of these systems presents a certain structure, which motivated the study of the implementation of Machine Learning algorithms to accelerate the solution process of these systems.

For this purpose, shallow artificial neural networks (ANN) were analyzed the present case \cite{merkel2018short}. This is because, the objective is to find light ANN structures that can emulate the solution of the system of algebraic equations obtained in the DGFEM discretization of transport problems at a reduced computational cost. Thus, the number of operations in the forward evaluation of the network must be minimized. Furthermore, we then applied a Dropout technique \cite{srivastava2014dropout} to the ANNs in order to reduce even further the number of operations required in the forward evaluation of the network.

In relation to radiation transport problems, neural networks has been scarcely used in radiation transport problems in comparison to other engineering disciplines. Some experience exist in using deep neural networks for building a parametric solar radiation spectrum \cite{ozturk2011artificial} and to improve the radiative scalar flux approximation in space radiation sensors \cite{bellotti1993comparison}. However, they have not been yet applied to accelerate the computationally expensive transport problems. Hence, in the present article, we demonstrate how to use neural networks for this purpose.

This work is organized as follows. In the next section, the DGFEM discretization of transport problems is analyzed in detail. Then, a review of the process leading to the design of the shallow ANNs used in this work is provided. Finally, the fitted ANNs are applied to four different test cases, which have been designed to incrementally test the performance of ANNs.


\section{DGFEM formulation of radiation transport problems}

The equation for radiation transport is an integro-differential equation that deals with the conservation of particles in space, angle, energy and time \cite{lewis1984computational}. Lets assume a domain in space $\mathbf{r}$, defined as $\mathcal{D}$, which is a spatial subdomain of $\mathbb{R}^d$ (with $d = 1,2,3$ depending on the dimension of the problem). Lets also assume that $\partial \mathcal{D}$ is its boundary that have an outward unit normal vector $\mathbf{n}$. We suppose that the angular direction of flight of particles can be represented by projecting this direction of flight on a references unit sphere, thus indicating the solid angle $\mathbf{\Omega}$ that this direction of flight makes in the sphere. This projection is generally made in $\mathcal{S}^2$ for 3D problems ($\mathcal{D} \subset \mathbb{R}^3$) and on a unit circle for 1D and 2D problems ($\mathcal{D} \subset \mathbb{R}^{1,2}$). Moreover, we assume a continuous representation of energy $E \in \mathbb{R}$ and time $t \in \mathbb{R}$. 

Following a probabilistic approach, the probability density function for a particle can be defined as $n = n(\mathbf{r}, \mathbf{\Omega}, E, t)$. To deal with reaction rates of particles that are streaming through a domain, it is more convenient to define the angular flux of particles as $\psi(\mathbf{r}, \mathbf{\Omega}, E, t) = vn(\mathbf{r}, \mathbf{\Omega}, E, t)$, where $v$ is the velocity of the streaming particle. Without loosing generality, we will assume that the problem is steady state and that particles are monoenergetic. For transient energy dependent problems, a time-marching scheme and a multigroup formulation can be applied for generalizing the results in the present article. In terms of the angular flux, the equation for conservation of particles states that for a differential element in the representative variables, the total loss of particles by streaming and absorption in the material is equal the number of particles produced by the sources in the material. This equation reads as follows:

\begin{equation}
\mathbf{\Omega} \bullet \nabla \psi(\mathbf{r}, \mathbf{\Omega}) + \sigma_t(\mathbf{r}) \psi(\mathbf{r}, \mathbf{\Omega})= Q(\mathbf{r}, \mathbf{\Omega})
\label{Eq:T1}
\end{equation}

Where the total interaction cross section $\sigma(\mathbf{r})$ has been defined, which specify the number of collisions per unit volume of the material once multiplied by its density and the angular flux. Moreover, the total source term comes from two different sources. First, particles that coming from an angle $\mathbf{\Omega}'$ interact with the atoms in the material and recoils into the angle  $\mathbf{\Omega}$. Second, due to the external production of particles in the material. In mathematical terms this reads as follows:

\begin{equation}
Q(\mathbf{r}, \mathbf{\Omega}) = \int_{4\pi} \sigma_s(\mathbf{r}, \mathbf{\Omega}' \rightarrow \mathbf{\Omega}) \psi(\mathbf{r}, \mathbf{\Omega}') d\mathbf{\Omega}' + Q_{ext}(\mathbf{r}, \mathbf{\Omega})
\label{Eq:T2}
\end{equation}

where we have defined the differential scattering cross section as $\sigma_s(\mathbf{r}, \mathbf{\Omega}' \rightarrow \mathbf{\Omega}$, which can be associated to the probability of a particle streaming in direction $\mathbf{\Omega}'$ to recoil into direction $\mathbf{\Omega}$. This differential scattering cross section should be a function of the cosine of the angle formed between the incident and outgoing directions, i.e. $\sigma_s(\mathbf{r}, \mathbf{\Omega}' \rightarrow \mathbf{\Omega}) =  \sigma_s(\mathbf{r}, \mathbf{\Omega}' \bullet \mathbf{\Omega})$. In order to be able to solve the problems numerically, it is customary to expand this cross section into Legendre polynomials as follows:

\begin{equation}
\sigma_s(\mathbf{r}, \mathbf{\Omega}' \rightarrow \mathbf{\Omega}) = \sum_{n=0}^N \frac{2n+1}{4 \pi} \sigma_{s,n}(\mathbf{r}) P_n(\mathbf{\Omega}' \bullet \mathbf{\Omega}) 
\label{Eq:T3}
\end{equation}

Since the Legendre polynomials satisfy the addition theorem \cite{cruzan1962translational}, the differential dependence of the scattering cross section can be expanded as follows:

\begin{equation}
P_n(\mathbf{\Omega}' \bullet \mathbf{\Omega}) = \frac{4 \pi}{2n+1} \sum_{m = -n}^n Y_{n,m}(\mathbf{\Omega}) Y^{*}_{n,m}(\mathbf{\Omega}')
\label{Eq:TLDec}
\end{equation}

where $Y_{n,m}$ are  spherical harmonics functions of polar angle order $n$ and azimuth angle order $m$ and $Y_{n,m}$ is its complex conjugate. This allows to get a separable expression for the differential cross section as follows:

\begin{equation}
\sigma_s(\mathbf{r}, \mathbf{\Omega}' \rightarrow \mathbf{\Omega}) = \sum_{n=0}^N \sum_{m=-n}^n \sigma_{s,n}(\mathbf{r}) Y_{n,m}(\mathbf{\Omega}) Y^{*}_{n,m}(\mathbf{\Omega}')
\label{Eq:T4}
\end{equation}

Applying this back into the source term \ref{Eq:T2} and injecting the source term into the transport equation, the following expression is obtained:

\begin{equation}
\mathbf{\Omega} \bullet \nabla \psi(\mathbf{r}, \mathbf{\Omega}) + \sigma_t(\mathbf{r}) \psi(\mathbf{r}, \mathbf{\Omega})= \sum_{n=0}^N \sum_{m=-n}^n \sigma_{s,n}(\mathbf{r}) Y_{n,m}(\mathbf{\Omega}) \phi_{n,m}(\mathbf{r}) + Q_{ext}(\mathbf{r}, \mathbf{\Omega})
\label{Eq:T5}
\end{equation}

\begin{equation}
\phi_{n,m}(\mathbf{r}) = \int_{4\pi} Y^{*}_{n,m}(\mathbf{\Omega}') \psi(\mathbf{r}, \mathbf{\Omega}') d\mathbf{\Omega}'
\label{Eq:T6}
\end{equation}

where the $(n,m)$ moments of the flux $\phi_{n,m}(\mathbf{r})$ have been defined in equation \ref{Eq:T6}. In general, an iterative process is carried out for the solution of equations \ref{Eq:T5} and \ref{Eq:T6} known as scattering source iteration for the solution of this system. In this one, $\psi(\mathbf{r}, \mathbf{\Omega})$ is solved from equation \ref{Eq:T5} assuming a known distribution of $\phi_{n,m}(\mathbf{r})$, this solution is injected back into equation \ref{Eq:T6}, the moments of the flux $\phi_{n,m}(\mathbf{r})$ are recomputed and injected back into equation \ref{Eq:T5} to carry on iterations.

For being able to deal with this equation computationally, a discretization needs to be performed both in angle ($\mathbf{\Omega}$) and space ($\mathbf{r}$). First, the S$_N$ is used for the angular discretization. This one consists in solving the angular flux in equation \ref{Eq:T5} on a set of quadrature directions in angle ($\mathbf{\Omega}_d$) that then allows approximating with good accuracy the integral in equation \ref{Eq:T6}. By introducing this approximation, the set of coupled equations \ref{Eq:T5}-\ref{Eq:T6} reads as:

\begin{equation}
\mathbf{\Omega}_d \bullet \nabla \psi(\mathbf{r}, \mathbf{\Omega}_d) + \sigma_t(\mathbf{r}) \psi(\mathbf{r}, \mathbf{\Omega}_d)= \sum_{n=0}^N \sum_{m=-n}^n \sigma_{s,n}(\mathbf{r}) Y_{n,m}(\mathbf{\Omega}_d) \phi_{n,m}(\mathbf{r}) + Q_{ext}(\mathbf{r}, \mathbf{\Omega}_d)
\label{Eq:T7}
\end{equation}

\begin{equation}
\phi_{n,m}(\mathbf{r}) \approx \sum_{d=1}^D Y^{*}_{n,m}(\mathbf{\Omega}'_d) \psi(\mathbf{r}, \mathbf{\Omega}'_d) \Delta \mathbf{\Omega}'_d
\label{Eq:T8}
\end{equation}

where the individual sets ($\mathbf{\Omega}_d, \Delta \mathbf{\Omega}_d$) represent the direction and the weight of the quadrature. Different quadrature sets may be choosen according the the type of particles and the characteristics of the material in the problem being solved. The hyperbolic problem is closed with the specification of the inlet fluxes in the boundaries of the domain as follows:

\begin{equation}
\psi(\mathbf{r}, \mathbf{\Omega}_d) = \psi_{inc}(\mathbf{r}, \mathbf{\Omega}_d), \qquad \mathbf{r} \in \partial{\mathcal{D}}, \mathbf{n} \bullet \mathbf{r} < 0
\label{Eq:TBC}
\end{equation}

In order to perform the space discretization, the Discontinuous Finite Element Method is used, as previously mentioned. In this one we assume that the space is discretized into a set of computational cells of volume $\mathcal{V}$ and boundary surface $\partial \mathcal{V}$. For compatibility with the angular discretization it results convenient to split the boundary of the cell in an inflow (+) and an outflow (-) boundary, i.e. $\partial \mathcal{V} = \partial \mathcal{V}_+ \cup \partial \mathcal{V}_-$. The positive part of the boundary is defined so that the incident direction have a negative projection over the normal to the cell boundary, i.e. $\mathbf{n}_{\partial \mathcal{V}} \bullet \mathbf{\Omega}_d < 0$; thus, radiation streaming in direction $\mathbf{\Omega}_d$ will be entering the cell $\mathcal{V}$ over the positive boundary $\partial \mathcal{V}_+$. Oppositely, the negative boundaries are defined such that $\mathbf{n}_{\partial \mathcal{V}} \bullet \mathbf{\Omega}_d > 0$ and, hence, radiation streaming in direction $\mathbf{\Omega}_d$ will be existing the $\mathcal{V}$ over the boundary $\partial \mathcal{V}_-$ \cite{wang2009high}. 

Once the domain have been identified, the next step is to propose a set of $J$ trail functions for representing the spatial dependence of the angular flux in any cell $\mathcal{V}_i$. This reads as follows:

\begin{equation}
\psi_i(\mathbf{r}, \mathbf{\Omega}_d) = \sum_{j=1}^J u_{ij}(\mathbf{r}) \psi_j(\mathbf{\Omega}_d)
\label{Eq:T9}
\end{equation}

Where $u_{ij}(\mathbf{r})$ is the $j$'th trial function for cell $i$. A similar decomposition can be performed for the moments of the scalar flux ($\phi_{n,m}$). The next step, as is classically done in finite elements, consists in adding degrees of liberty in the representation by multiplying with a set of $K$ weight function $w_{ik}(\mathbf{r})$ for cell $i$ and integrating over the volume of cell $\mathcal{V}_i$. By performing these operations in equation \ref{Eq:T5} and introducing the trial representation of the angular flux \ref{Eq:T9} the following terms are obtained:

Streaming:

\begin{centering}
$\sum_{j=1}^J \sum_{k=1}^K \int_{\mathcal{V}_i} w_{ik}(\mathbf{r}) (\mathbf{\Omega}_d \bullet \nabla u_{ij}(\mathbf{r})) \psi_j(\mathbf{\Omega}_d) dV =$ \\
$\sum_{j=1}^J \sum_{k=1}^K \int_{\partial \mathcal{V}_{+,i}} w_{ik}(\mathbf{r}) u_{ij}(\mathbf{r}) (\mathbf{\Omega}_d \bullet \mathbf{n}_{\partial \mathcal{V}_+}) \psi_j(\mathbf{\Omega}_d) dV +
\sum_{j=1}^J \sum_{k=1}^K \int_{\partial \mathcal{V}_{-,i}} w_{ik}(\mathbf{r}) u_{ij}(\mathbf{r}) (\mathbf{\Omega}_d \bullet \mathbf{n}_{\partial \mathcal{V}_-}) \psi_j(\mathbf{\Omega}_d) dV + $\\
$- \sum_{j=1}^J \sum_{k=1}^K \int_{\mathcal{V}} u_{ij}(\mathbf{r}) \psi_j(\mathbf{\Omega}_d) (\mathbf{\Omega}_d \bullet \nabla w_{ik}(\mathbf{r}) dV = $\\
$(\mathbf{\Omega}_d \bullet \mathbf{n}_{\partial \mathcal{V}_+})\mathbf{L}^+ \mathbf{\psi}_d^{+} + (\mathbf{\Omega}_d \bullet \mathbf{n}_{\partial \mathcal{V}_-})\mathbf{L}^-  \mathbf{\psi}_d^{-} + \mathbf{\Omega} \bullet \mathbf{L} \mathbf{\psi}_d$

\end{centering}

Total interaction:
$
\sum_{j=1}^J \sum_{k=1}^K \int_{\mathcal{V}_i} \sigma_{ti} w_{ik}(\mathbf{r}) u_{ij}(\mathbf{r}) \psi_j(\mathbf{\Omega}_d) dV = 
\mathbf{M} \mathbf{\psi}_d
$
\\

Scattering:
$
\sum_{n=0}^N \sum_{m=-n}^n \sum_{j=1}^J \sum_{k=1}^K \int_{\mathcal{V}_i} \sigma_{s,n}(\mathbf{r}) Y_{n,m}(\mathbf{\Omega}_d) \phi_{n,m} w_{ik}(\mathbf{r}) u_{ij}(\mathbf{r}) dV = 
\mathbf{M} [\mathcal{M} \mathbf{\sigma}_s \mathbf{\phi}]
$

Source term:
$\sum_{j=1}^J \sum_{k=1}^K \int_{\mathcal{V}_i} w_{ik}(\mathbf{r}) Q_{ext}(\mathbf{r}, \mathbf{\Omega}_d)dV = 
\mathbf{Q}_d$

where Green's theorem has been used in the streaming term. Furthermore, the upwind streaming matrix $\mathbf{L}^+$, the downwind streaming one $\mathbf{L}^-$, the gradient matrix $\mathbf{L}$, the mass matrix $\mathbf{M}$, and the external source term vector $\mathbf{Q}_d$ have been introduced. Moreover, equation \ref{Eq:T8} can be put in matrix form as $\mathbf{\phi} = \mathcal{D} \mathbf{\psi}_d$. Note that the shape and size of these matrices and vector, will depend in the order and shape of the trial and test functions. In the present case, the results presented were computing using $P_1$ Galerkin Finite Elements. The final system of equations reads as follows:

\begin{equation}
(\mathbf{\Omega}_d \bullet \mathbf{n}_{\partial \mathcal{V}_+})\mathbf{L}^+ \mathbf{\psi}_d^{+} + (\mathbf{\Omega}_d \bullet \mathbf{n}_{\partial \mathcal{V}_-})\mathbf{L}^-  \mathbf{\psi}_d^{-} + \mathbf{\Omega})d \bullet \mathbf{L} \mathbf{\psi}_d + \mathbf{M} \mathbf{\psi}_d = \mathbf{M} [\mathcal{M} \mathbf{\sigma}_s \mathbf{\phi}] + \mathbf{Q}_d,
\label{Eq:T10}
\end{equation}

\begin{equation}
\mathbf{\phi} = \mathcal{D} \mathbf{\psi}_d.
\label{Eq:T11}
\end{equation}

These system of equations have to be solved for each cell $\mathcal{V}_i$ in the domain. However, the fluxes in the inflow faces and the ones in the outflow faces still need to be determined this system of equations. Interpolations of different orders can be performed to find these values based in spatial moment balances of equations \ref{Eq:T10} and \ref{Eq:T11} \cite{vikas2013radiation}. However, in the present case, we limit ourselves to first order interpolations \cite{plimpton2000parallel}. In these ones, the inlet fluxes $\mathbf{\psi}_d^{-}$ are taken as the flow in shared faces with upwind neighbours of the cell. Moreover, the outlet fluxes $\mathbf{\psi}_d^{+}$ are taken as the fluxes predicted by the current solution in the face in the outlet surfaces. In other words, for each cell, a system of the following form is being solved:

\begin{equation}
  \begin{split}
   \mathbf{A}_{\psi} \mathbf{\psi}_d = \mathbf{b}_{\psi} \\
   \mathbf{\phi} = \mathcal{D} \mathbf{\psi}_d 
   \end{split}
\label{Eq:T12}
\end{equation}

where we have defined $\mathbf{A}_{\psi} = (\mathbf{\Omega}_d \bullet \mathbf{n}_{\partial \mathcal{V}_-})\mathbf{L}^- + \mathbf{\Omega} \bullet \mathbf{L} + \mathbf{M} \mathbf{\psi}_d$ and $\mathbf{b}_{\psi} = -(\mathbf{\Omega}_d \bullet \mathbf{n}_{\partial \mathcal{V}_+})\mathbf{L}^+ + \mathbf{M} [\mathcal{M} \mathbf{\sigma}_s \mathbf{\phi}] + \mathbf{Q}_d$. The solution process for this system \ref{Eq:T12} for a direction $\mathbf{\Omega}_d$ starts from the cells next the boundaries were the inflow angular fluxes are known for this direction. Once these cells are solved, information on the upwind fluxes becomes available to its neighbour cells and the problem can be solved in these ones. This solution process is known as sweeping because a sweeping is performed through the domain for finding the solution. The sweeping process requires a large degree of scheduling in order to make it performance in massively parallel computer architectures. Once the angular fluxes have been computed for the whole domain, the scalar fluxes $\mathbf{\phi}$ are updated for all cells in the domain, according to the second equation in the system \ref{Eq:T12}, which in turn updates the right hand side vector ($\mathbf{b}_{\psi}$) in the first equation \cite{adams2002fast}. Iterations are performed until the errors in the scalar flux are smaller than a fixed tolerance.

The most time consuming step during the solution process, consists in the inversion of the first equation in \ref{Eq:T12}. These systems are not large (e.g. for hexahedral cells with P1 finite elements systems of size $(8 \times 8)$ are obtained) and, thus, are generally solved by direct Gauss Elimination. However, they have to be solved for each cell in the domain and for every sweeping direction. For example, for a domain discretized in space with $10^5$ cells and in angles with $10^3$ sweeping direction, it means that $10^8$ systems of the form $\mathbf{A}_{\psi} \mathbf{\psi}_d = \mathbf{b}_{\psi}$ have to be solved per sweeping direction.

Nevertheless, as seen in the definition of the system \ref{Eq:T12}, there is an analogous structure repeating when solving the system in each cell and direction. 

For instance, when changing the sweeping direction, the weight of the gradient ($\mathbf{L}$) and downwind ($\mathbf{L}^-$) matrices will change in the matrix of coefficients $A_{\psi}$, but their structure will be unchanged. Moreover, when changing the sweeping direction, the influence of the upwind matrix ($\mathbf{L}^+$) will change in the right hand vector, but the way in which this vector changes will be uniform in space by $\mathbf{L}^+$. 

In a similar manner, changing the shape of the cells will change the construction of the upwind, downwind, gradient and mass matrix, but it will do it in a proportional way for all of them as observed in the integral definition of these matrices. 

Finally, not that the right hand side vector is also composed by the scattering and external source. The external source is fixed for every-cell during source iterations and, generally, it is considered to be constant within the cell for a sweeping direction. Hence, it only add a constant additive factor to the right hand side. Moreover, the scattering term, add a source that is proportional to the scalar flux level within the cell, which could already be predicted by the previous solution obtained for the angular flux.

The objective of this work was to exploit the structure observed in system \ref{Eq:T12} in order to develop a faster way to solve equation for the angular fluxes. For this purpose, different Machine Learning techniques were studied, including kernel regression, support vector machines, self-organized mapping methods, and Artificial Neural Networks (ANNs). Of these techniques, it was found that ANNs provided the best ratio between computational performance and fitting accuracy. For this purpose, ANNs are presented in further detail in the next section.


\section{An artificial neural network as a predictor for DGFEM solutions}

Artificial Neural Networks (ANNs) has proven to be highly efficient for modeling non-linear processes without a-priori specified relationships between the input and the output variables \cite{chen2010robust}. In particular, there is a growing interest in using these ANNs for fitting highly nonlinear processes in continuum mechanics \cite{tracey2015machine}, as well as for model discovery \cite{ghahramani2015probabilistic}. In the present case these ANNs will be used for being able to determine the solutions for the angular fluxes in system \ref{Eq:T12}, without the need of inverting the matrix of coefficients.

The first approach developed consisted on trying to train a shallow neural networks as a replacement for Gaussian Elimination. For this purpose, the coefficients of $A_{\psi} \in \mathbb{R}^{64}$ and the right-hand-side vector $b_{\psi}$ in equation \ref{Eq:T12} were feed as inputs to the neural networks, building an input vector $\mathcal{I} \in \mathbb{R}^{72}$. The output vector consisted on the angular fluxes $\mathcal{O} \in \mathbb{R}^8$. However, the number of connections that should be evaluated by this network for a simple one layer case is already 640. Even though some optimization techniques could be envisioned, the cost of a feed-forward evaluation of this network is larger than the cost of doing Gauss Elimination in \ref{Eq:T12}. Hence, this approach was abandoned.

\begin{figure}
  \centering
  \includegraphics[scale=0.4]{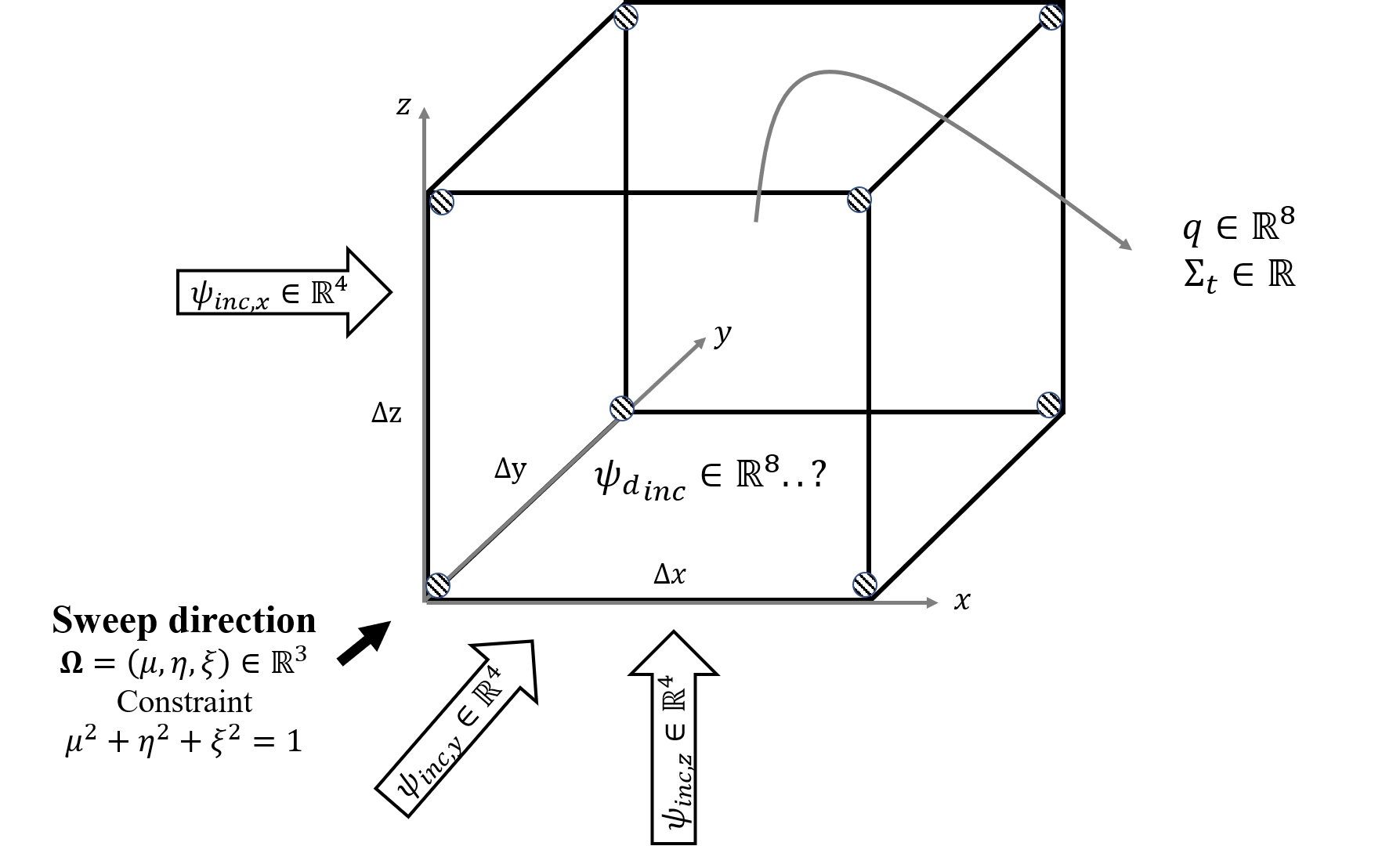}
  \caption{Sketch of the DFEM transport problem in an hexahedral cell.}
  \label{fig:onecellprob}
\end{figure}

In order to design a better approach, the number of inputs in the ANN have to be reduced. For this purpose, the general transport problem for one cell was re-examined as solved in the present case. An example sketch of the problem for an hexaheral cell is shown in Figure~\ref{fig:onecellprob}. The independent input values are:
\begin{itemize}
    \item 4 input values from each of the 3 inflow faces ($\psi_x^{inc} \in \mathbb{R}^4, \psi_y^{inc} \in \mathbb{R}^4, \psi_z^{inc} \in \mathbb{R}^4$),
    \item 2 independent angles that determine the sweeping direction $(\mu, \eta) \in [-1,1] \subset \mathbb{R}^2$,
    \item 8 input values determine the scattering and external source in the cell $q \in \mathbb{R}^8$,
    \item 1 input value determine the total cross section in the domain $\sigma_t \in \mathbb{R}^+ \subset \mathbb{R}$
    \item 3 input values determine the shape of the cell $(\Delta x, \Delta t, \Delta z) \in \mathbb{R}^{+,3} \subset \mathbb{R}^3$
\end{itemize}
Hence, generalizing, an input vector in $\mathcal{I} \in \mathbb{R}^26$ fully determine the solution in the one cell problem. The desired output vector are still the values of the angular fluxes in each of the 8 nodes used for the linear finite element discretization in the hexahedral cell $\mathcal{O} \in \mathbb{R}^8$. Note that for a fully connected one layer network, the total number of connections is of 272, which provides a promising path for optimizing the number of operations performed with Gauss Elimination.

\begin{figure}
  \centering
  \includegraphics[scale=0.6]{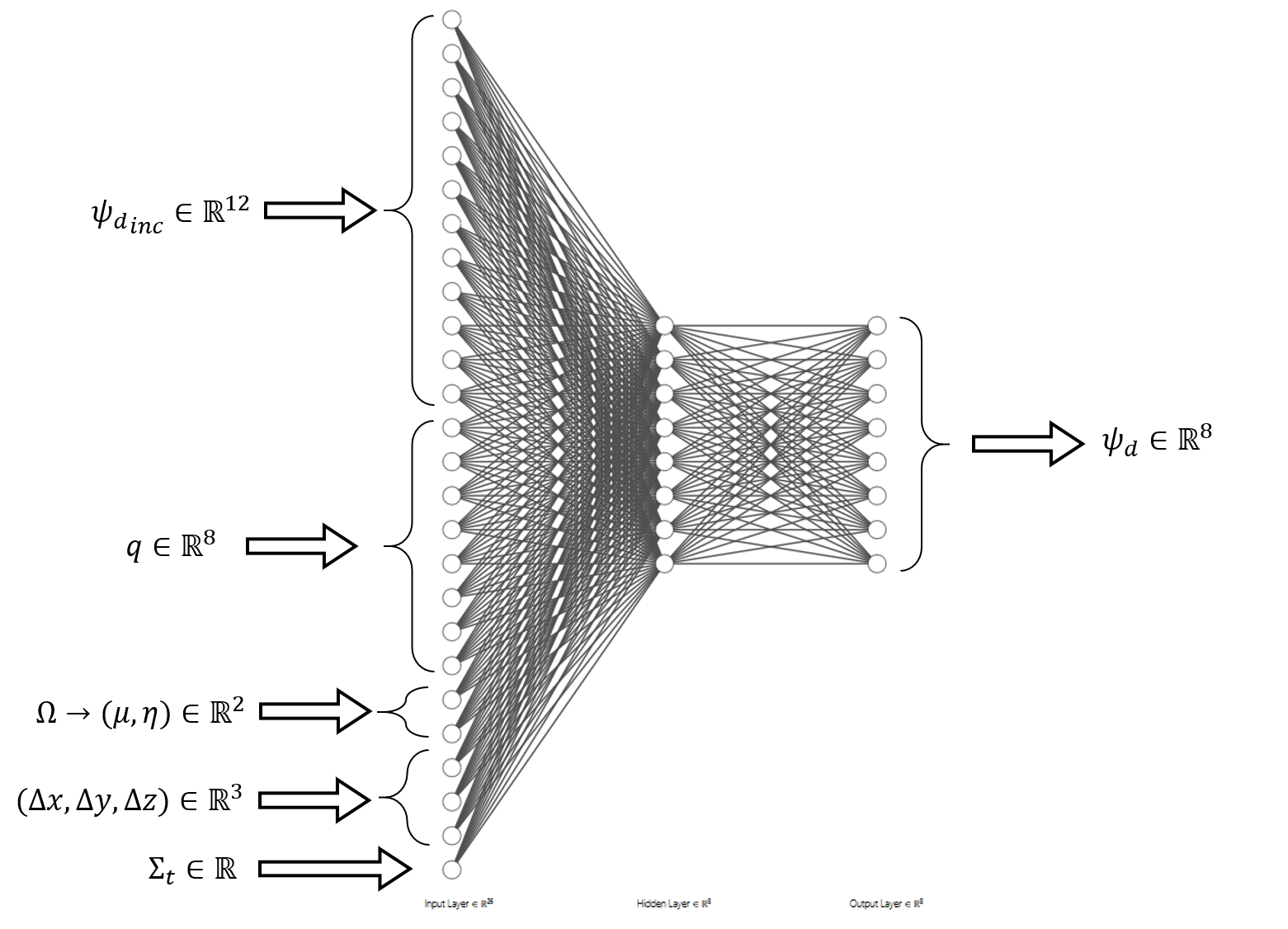}
  \caption{Example of the feed-forward ANNs proposed for the present case.}
  \label{fig:ANNConcept}
\end{figure}

A simple multi-layer, feed-forward neural network \cite{glorot2010understanding} was proposed to be used as the starting point in this case. An example configuration of a densely connected one layer network used in this case is shown in Figure~\ref{fig:ANNConcept}. The elements in charge of the filtering the information are called neurons. These neurons are arranges in layers. The first layer is called the input layer. In this case, this layer takes the inputs vector $\mathcal{I} \in \mathbb{R}^{26}$. The output comes from the last layer of this network, which in this case is $\mathcal{O} \in \mathbb{R}^8$. The layer(s) in between these two are called hidden layer(s). With fixed input and output vectors, the model complexity can be increased by increasing the number of hidden layers or increasing the number of neurons in this hidden layers.

In this analysis, we have the non-typical constraint of wanting rapid feed-forward evaluations of the ANN. 

For any given number of input nodes, the model complexity may be increased by increasing the hidden layers and/or increasing the number of nodes in the hidden layer(s). For the purposes of this analysis we used three-layered networks only. Each node is connected to all other nodes of adjacent layers and receives input and sends output through these connections. In the hidden layers the nodes simply sum the weighted values of the input variables, adds a bias term, and passes the value through a scaling function (or activation function in the neural net jargon). A three–layered neural network may mathematically be represented by

\begin{equation}
    \begin{split}
    h_k = f_k [\sum_{j=1}^n(w_{jk} \mathcal{I}_j + b_k)], \hspace{8pt} k = 1,...,26 \\
    \mathcal{O}_t = \theta_t [\sum_{k=1}^m(w_{kt} h_k + b_t)], \hspace{8pt} k = 1,...,8
    \end{split}
    \label{ANN:Eq1}
\end{equation}

where $h_k$ is the hidden–layer output, which is scaled by the activation function $f(x)$ defined below. The first subscript in the weights, $w$, indexes the nodes in the preceding layer and the second indexes the nodes in the current layer, $\mathcal{I}_j$, where $j = 1, ..., 26$ are the input values of the input layer, $b$ is the bias term, and $n$ is the number of the input–layer nodes. The hidden–layer output becomes the input values for the output layer, $o_t$, which uses the output scaling function $\theta(x)$ (defined below) to scale the output values to appropriate range for use in the transport problem. 

Different popular choices exist for the activation function, regarding the application for which they are implemented \cite{jain1996artificial}. Two popular functions for example are the linear rectifiers (ReLU) or sigmoid activation functions. In the present case, different popular choices were implemented. However, afterwards, we found convenient to tailor specific activation functions for our desired applications. 

For this purpose, two critical aspects were remarked. First, for incident fluxes into the cell, the fluxes obtained as solution should show a behaviour $\propto e^{-\sigma_t s}$, where $s$ is the distance traveled by a radiation ray within the cell. Furthermore, for the internal sources, this solution should behave as $\propto (1-e^{-\sigma_t s})$. However, the transport problem is linear on its inputs and sources, so a non-linear combination of inputs and sources in not desired. Moreover, the DFEM formulation that we have implemented can predict negative angular fluxes for some of the angular flux wights within cell $\mathcal{V}_i$, i.e. $\psi_{d,i,j} \not\in \mathbb{R}^+$. For this reason, we need a support for negative input that does not introduce an artificial aliasing in the solution.

Considering the previous requirements, a two step solution was developed. First, each of the inputs where linearly scaled to the expected range $[-1,+1]$, by modifying each of the inputs individually. For this purpose the maximum expected value $\mathcal{I}_j^{MAX}$ and the minimum expected one $\mathcal{I}_j^{MIN}$ were identified a-priori for each input $j$ and each input was rescaled as follows:

\begin{equation}
    {\mathcal{I}_{n,j}} = \frac{I_j - \alpha_j ||\mathcal{I}_j^{MIN},\mathcal{I}_j^{MAX}||}{|\mathcal{I}_j^{MAX} - \mathcal{I}_j^{MIN}|}
    \label{ANN:Eq2}
\end{equation}

where $||.||$ represents a norm of the input distribution, taken as the means between $\mathcal{I}_j^{MIN}$ and $\mathcal{I}_j^{MAX}$ in this case. Moreover, $\alpha_j$ was a parameter that was adjusted to maximize the linear behaviour of the solution after the hidden layer, in the present case we have used $\alpha_j = 0.27$ for all cases presented.

The second step, was to give an exponential behaviour by providing a filtering function that allowed to give exponential behaviour to with respect to the total cross section, but also allowed to maintain the linear behaviour on the fluxes and the sources of the transport problem. For this purpose, an hyperbolic tangent was proposed as activation function for the hidden layer. This one is:

\begin{equation}
    f(x) = tanh(x)
    \label{ANN:Eq3}
\end{equation}

Observe that in the limit as $x \rightarrow 0$ the hyperbolic tangent can be approximated by $tanh(x) \approx x$, which allows to retain a linear behaviour near zero. Moreover, as we increase or decrease the values in $x$ we have $tanh(\pm x) \propto e^{\pm x}$. Hence, we expect that the incident fluxes and the sources will be mapped into the linear regions of $f(x)$ by the wights $w_{jk}$ and that the weights associated to the total cross section will help map the net flux and source term in the cell towards the region were the activation function behaves exponentially. This behaviour was observed in practice after optimizing the ANNs.

To be able to use the neural network output directly as the solutions for the angular flux, we have allowed the output values to be scaled by the actual values that they show in the solution. For this purpose, exponential linear units (ELUs) \cite{clevert2015fast} are used for  the output layer. This means that the functions $\theta_t(x)$ were defined by:

\begin{equation}
    \theta_t(x) =
    \begin{cases} x, & \mbox{if } x>=0 \\ \beta(e^x-1), & \mbox{if } x<0 \end{cases}
    \label{ANN:Eq4}
\end{equation}

The advantage of this function is that the positive outputs can be linearly scaled with the weights of the ANN, thus, extending the support for the output of the net to $\mathbb{R}^+$ for positive solutions. However, it filters the negative solutions for the angular fluxes to a maximum value of $-\beta$ and, hence, it reduces the influence of large negative solutions that may be obtained for the angular fluxes in the DFEM discretization. In short, the output of the ANN is bounded to $\mathcal{O} \in (-\beta,+\infty)$. The value of $\beta$ is then optimized to improve the performance of the net under negative solution, without affecting significantly the mechanics of the transport process in the domain. In the present case a value of $\beta = 0.1 \mbox{max}(\psi_d^{inc})$ was observed to yield good results.

Once the ANN has been formulated, the next step was studying the sensitivity of the ANN to the set of hidden layers and number of neurons in this hidden layer. For this purpose we build a dataset with $10^5$ different one-cell problems. In this problems, random values are imposed for the input values of the ANN and solutions for the angular fluxes are obtained. Then, we select 90\% of these problems in the training set of the ANN, i.e. they are used in the loss function to fit the weights in the connections of the ANN by back-propagation. Furthermore, 10\% of this problems are used in the testing set, i.e. for evaluating the performance of the ANN in problems that have not been specifically used for training. 

\begin{figure}
  \centering
  \includegraphics[scale=0.6]{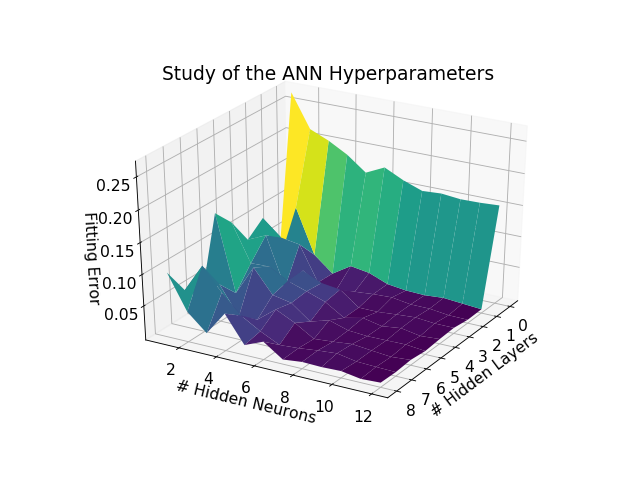}
  \caption{Study of the mean quadratic error over $10^4$ random testing problems for the ANN proposed as a function of the number of hidden layers and the number of neurons in the hidden layer.}
  \label{fig:MQE}
\end{figure}

The results obtained for the above mentioned study are presented in Figure~\ref{fig:MQE}. It is observed that the mean quadratic error over the training set is reduced as the number of hidden layers and the number of neurons in this hidden layer augment. However, it is also observed that for 8 or more neurons in the hidden layer, the error rapidly decreases when adding one hidden layer and remains approximately constant as more hidden layers are added. Since the objective in the present case is to reduce the number of operations in the feed-forward evaluation of the ANN, an ANN with 1 hidden layer and 8 neurons in the hidden layer, as the one shown in Figure~\ref{fig:ANNConcept},is selected.

Furthermore, in order to reduce even more the number of operations required in the feed-forward evaluation of the ANN, a Dropout technique is implemented \cite{srivastava2014dropout}. This one consists of an implicit regularization technique. In this one, the connection in the ANN with the weakest weight is dropped, the network is retrained, and a the mean square error of the network is analyzed. If the performance was not affected by dropping this connection, then the new weakest connection if drop, in the opposite case, the process stops. This process is performed in the proposed ANN until the mean square error over the training set deteriorates in more than 0.1\%.

\begin{figure}
  \centering
  \includegraphics[scale=0.6]{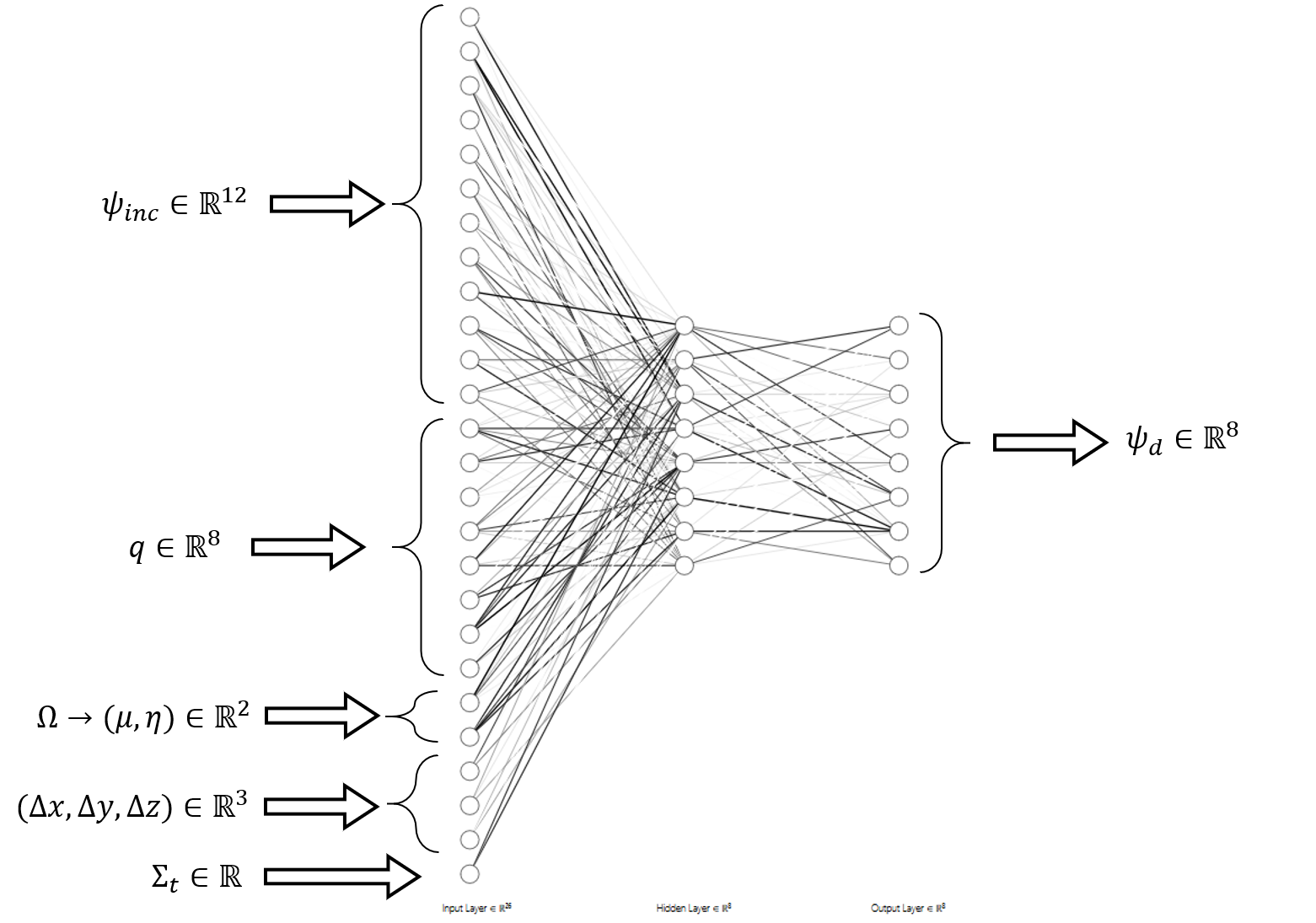}
  \caption{Schematic representation of the ANN after the dropout technique has been implemented.}
  \label{fig:DO}
\end{figure}

A schematic representation of the final ANN obtained after the dropout technique has been implemented is presented in Figure~\ref{fig:DO}. In this one, the dropped out connections have been colored in white. The final ANN with has 16\% less connections and, hence, a feed-forward evaluation of the ANN can be performed ~10\% faster than a fully connected ANN.

\begin{figure}
  \centering
  \includegraphics[scale=0.6]{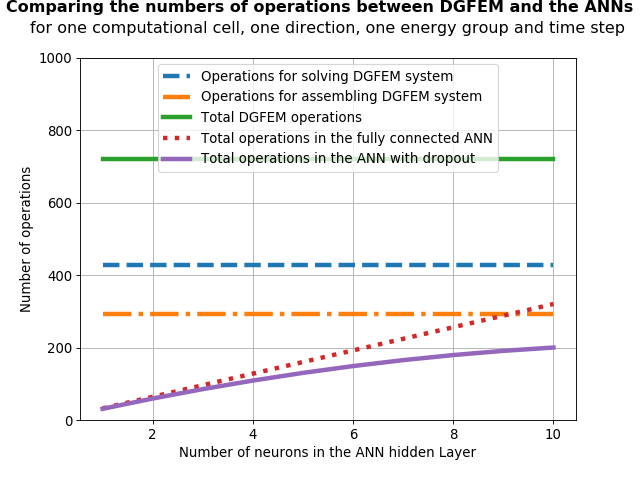}
  \caption{Comparison of the number of computational operations required for assembling and solving the fluxes with Gauss eliminations vs the number of computational operations required to obtain the solution with the present ANN configuration.}
  \label{fig:CompANNGauss}
\end{figure}

As a final comparison in this section, the number of computational operations required for the assembly and Gauss elimination of the system in the Equation~\ref{Eq:T12} is compared against the one required for a feed-forward evaluation of the ANN in Figure~\ref{fig:CompANNGauss}. The Figure presents the results as a function of the total number of neurons in the hidden layer. Two main conclusions can be driven from this study. First, the influence of dropout causes the ANN to maintain an approximately constant number of connections even when more than 8 neurons in the hidden layer are added, supporting the current selection of 8 neurons in the hidden layer. Most importantly, the ANN reduces the the number of operations required to compute the angular fluxes by ~75.6\%.

The results obtained in the present section show that a big acceleration may be possible when solving transport problems, while introducing only small errors ($<0.1\%$) for these problems. Nevertheless, this bounded small errors may not be maintained when solving transport problems in bigger domains, where the solution of the problem in one cell is coupled to its neighbours. Thus, in the next section we examine the performance of the net for four 3D test cases.




\section{Evaluating the performance of ANNs in four test cases}

In the present case, the performance of the the developed ANN in evaluated against the classical solution obtained with assembling system \ref{Eq:T12} and solving it by Gauss Elimination. Four different test cases are proposed to test the performance of the ANN architecture in incrementally more complex problems. These problems are:
\begin{enumerate}
    \item Problem 1: incident flux with no internal sources in a thick purely absorbing medium
    \item Problem 2: incident flux with no internal sources in a quasi-transparent medium
    \item Problem 3: incident flux with internal sources in a thick purely absorbing medium
    \item Problem 4: incident flux with internal sources in a thick medium with a large scattering ratio
\end{enumerate}

In all this cases the domain $\mathcal{D}$ consists of a cube that is discretized with hexahedral cells. Different space and angle discretizations in this cube are evaluated for each problem to assess their influence in the performance of the ANN. The solutions by ANNs is segmented by direction. This means that 8 different ANNs will be developed for the 8 different sweeping octants in the cube (each octant direction goes with the vertices of the cube). This segmented strategies allowed to improve the performance of the ANNs over the test problems by ~4-7\%.

The training of the ANNs in each direction is performed on a set of $9\times10^4$ solutions to one-cell problems. These problem were generated by performing a uniform random sampling on the incident fluxes, internal sources, total cross section, sweeping direction and shape of the cell and retrieving the solutions for the angular fluxes in the hexahedral cells. The testing set for the ANN consists of $1\times10^4$ test problems produced by following an equivalent methodology than the one used for the training set. The error function is defined as the mean absolute lost over the training set with a regularizer that induces the optimization algorithm to obtain the homogeneous solutions as follows:

\begin{equation}
    \epsilon = \frac{1}{N} \sum_{j=1}^N \frac{1}{8} \sum_{d=1}^8 |\psi_{d,j} - \psi^{ANN}_d(\mathcal{I}_j)| + \frac{\lambda}{N} \sum_{j=1}^N \frac{|\sum_{d=1}^8 \psi_{d,j} - \psi^{ANN}_d(\mathcal{I}_j)|}{\gamma_1 |\sum_{d=1}^N \psi_{d,j}|+ \gamma_2}.
    \label{Test:Eq1}
\end{equation}

The first term in the RHS of equation is simply the mean absolute difference between the solutions obtained by Gauss Elimination $\psi_{d,j}$ and the solutions predicted by the ANN $\psi^{ANN}_d(\mathcal{I}_j)$ for all the training problems ($N = 9\times10^4$). The second parameter, is a regularization parameter, which objective is to increase the penalty of the absolute error if the solution is closer to zero. This parameter was crucial in the present case to induce the ANN training to preserve the homogeneous solution for the fluxes, thus avoiding the build up errors during the streaming process of radiation. In the present test cases, the calibration parameters of the regularizer were tuned as $\lambda = 0.2$, $\gamma_1 = 10$ and $\gamma_2 = 0.01$.

The training of weights in the connections ANNs was performed by back-propagation following the definition of the error function \ref{Test:Eq1}. Furthermore, a constrain function was implemented to avoid evaluating the connections that have been discarded during dropout, when performing the optimization process. The Adam algorithm was used for training purposes. Given the low dimensionality of the set of weights in the ANN with respect to the number of training samples, smaller weights than the recommended arguments were used in the Adam algorithm, i.e. learning rates $lr \in [5\times10^{-5}, 2\times10^{-4}]$. These weights were selected for minimizing the error over the testing set. A total of $5\times10^4$ iterations of the Adam algorithm (epochs) were performed until the convergence of the weights. The mean absolute errors over the training set for the ANNs trained were in $[3\times10^{-5}, 11\times10^{-5}]$ and the maximum absolute error for each of these ANNs was $[2\times10^{-3}, 3\times10^{-3}]$. Moreover, the mean squared errors over the testing set were in $[4\times10^{-5}, 2\times10^{-4}]$ and the maximum absolute errors for each ANN were in $[8\times10^{-3}, 14\times10^{-3}]$.

Once the ANNs have been trained, they were replaced into a 3D DGFEM solver. In this sense, the modified solver works similarly than the classical solver, with the exception that the assembling and solution of system \ref{Eq:T12} is replaced by the construction of the input vector ($\mathcal{I}$) and a feed-forward evaluation of the ANN. The results obtained for the four test problems proposed are shown in the following subsections.

\subsection{Problem 1: Incident flux in a strong absorbing medium}

\begin{figure}
  \centering
  \includegraphics[scale=0.4]{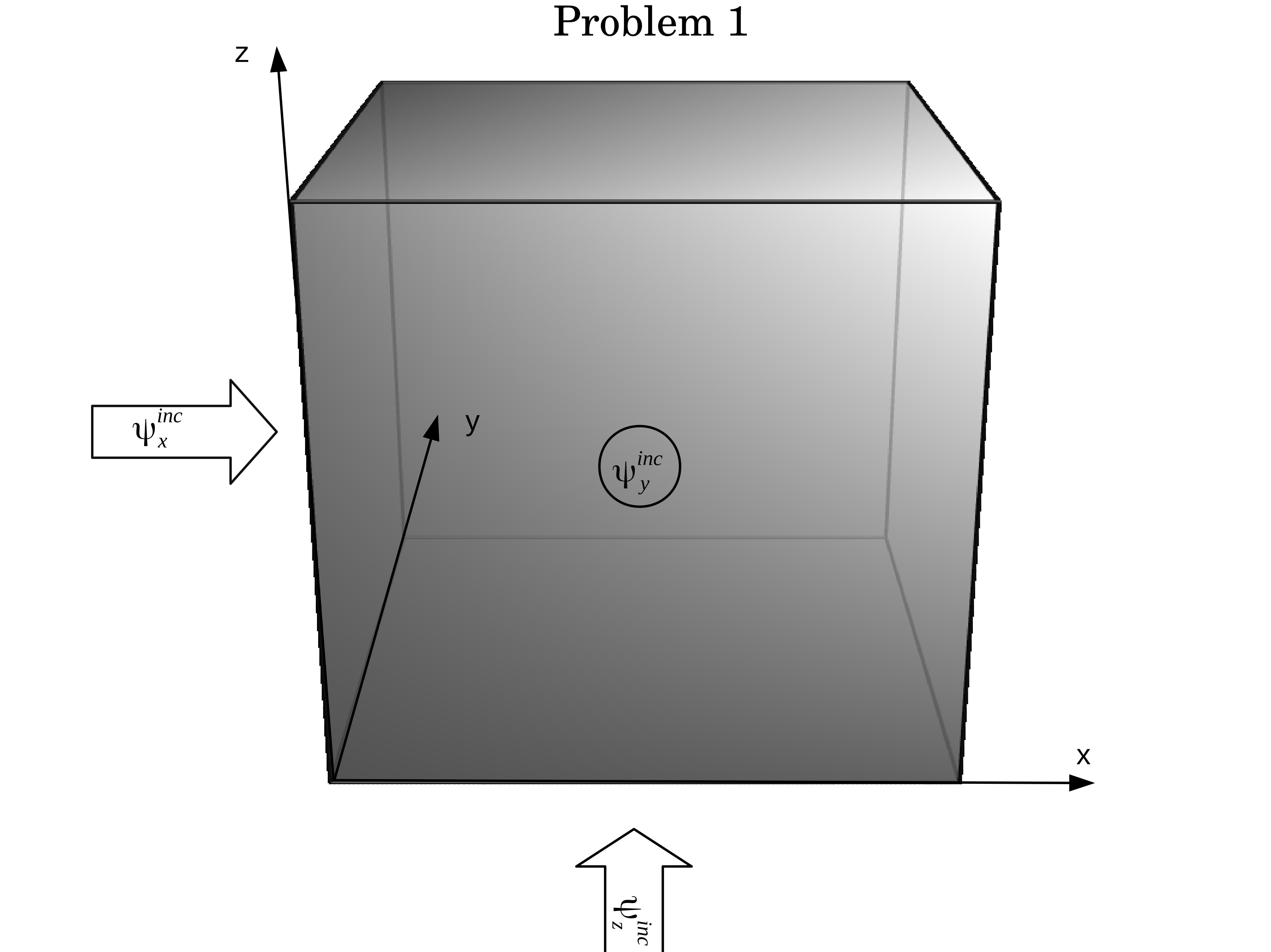}
  \caption{Problem 1.}
  \label{fig:fig1}
\end{figure}

As previously mentioned, Problem 1 consists of an incident flux on a thick purely absorbing medium. The parameters for Problem 1 are shown in Table \ref{tab:tableP1}. A sketch of the problem is shown in Figure~\ref{fig:fig1}. Incident fluxes of $(\psi_x^{inc}, \psi_y^{inc}, \psi_z^{inc})$ are imposed in the left, back, and bottom faces respectively. Void boundary conditions are imposed in the opposite right, front, and top faces. There are no internal sources in the domain. The total size of the domain for this problem is between $[100, 173]$ mean free paths. Hence, it is expected that the incident fluxes will rapidly reduce while streaming to the outflow faces. Since there is no scattering in this problem, no source iterations are performed and the problem converges one all cells have been swept for all directions.

In order to compare the solutions obtained with Gauss Elimination and with the ANN the mean absolute error over the domain is defined by:

\begin{equation}
    \epsilon_P = \frac{1}{N_x N_y N_z} \sum_{i=1}^{N_x} \sum_{j=1}^{N_y} \sum_{k=1}^{N_z} \frac{|\phi(x_i,y_i,z_i) - \phi^{ANN}(x_i,y_i,z_i)|}{|\phi(x_i,y_i,z_i)| + m}
    \label{Test:Eq2}
\end{equation}

where $\phi(x_i,y_i,z_i)$ is the scalar flux obtained with Gauss Elimination in cell $(i,j,k)$ and $\phi^{ANN}(x_i,y_i,z_i)$ the one obtained with the ANN. The parameter $m$ is introduced for avoiding divisions by zero and is set to $m = 10^{-8}$.

\begin{table}
 \caption{Parameters of Problem 1}
  \centering
  \begin{tabular}{llllllllllll}
    \toprule
    \multicolumn{3}{c}{Incident Fluxes $[n/{cm^2s}]$} & \multicolumn{3}{c}{Medium Properties} & \multicolumn{3}{c}{Lengths $[cm]$}  & \multicolumn{3}{c}{Discretization}  \\
    \cmidrule(r){1-3}
    \cmidrule(r){4-6}
    \cmidrule(r){7-9}
    \cmidrule(r){10-12}
    $\psi^{inc}_x$ & $\psi^{inc}_y$ & $\psi^{inc}_z$ &  $\Sigma_{TOT} [1/cm]$ & $\Sigma_S [1/cm]$ & $q [n/(cm^3s)]$ & $L_x$ & $L_y$ & $L_z$ & $N_x$ & $N_y$ & $N_z$  \\
    \midrule
    1.0 &  1.0 &  1.0 &  10.0 &  0 &  0 &  10.0 &  10.0 &  10.0 &  10-100 &  10-100 &  10-100 \\
    \bottomrule
  \end{tabular}
  \label{tab:tableP1}
\end{table}

The first test, consists of changing the discretization in angle of the problem, while keeping the spatial discretization fixes with $(\Delta x, \Delta y, \Delta z) = (0.1, 0.1, 0.1)$cm. The results obtained for an angular discretization using using $S_2$, $S_4$, $S_6$ and $S_8$ are presented in Table \ref{tab:tableResP1S}. It is observed that for $S_2$ the errors obtained are scaled by one order of magnitude with respect to the mean errors obtained in the testing set. This is because, despite the regularizer introduced introduced in the cost function \ref{Test:Eq1}, there is still a small monotonous behaviour in the errors near zero that causes a build up in the errors as radiation propagates downstream. 

To explain this better, let's suppose that a small $\delta$ error is produced in a cell when solving the radiation transport problem with the ANN in one cell. Then a flux of $\psi_d + \delta$ will be input as the upstream flux to the neighbours. Since the problem is linear and the behaviour of the error in the ANN is monotonous, it is expected that the neighbours will produce an error on the flux of $2*\delta$. Moreover, the neighbours of the neighbours will in turn produce errors of $3*\delta$, and so on. Nonetheless, this error build up is generally small because of the regularizer developed.

It can be further observed in Table~\ref{tab:tableResP1S} that the performance of the ANN is deteriorated as the number of sweeping direction per octant are increased. This is logical since the variance in the input vector of the ANN is increased and, hence, the variance of error in the expected output should also be increased. However, it is noted that this phenomenon approximately stabilizes for $N \geq 6$. The mean errors obtained in these conditions are of less than 2\%, which is completely acceptable for the present requirements. 

\begin{table}
 \caption{Error and acceleration generated by replacing Gauss Elimination with an ANN in Problem 1 for a mesh of $100 \time 100 \times 100$ cells}
  \centering
  \begin{tabular}{lll}
    \toprule
    \multicolumn{3}{c}{Results Problem 1}  \\
    \cmidrule(r){1-3}
    Order $S_N$ & Error ANN & Acceleration  \\
    \midrule
    2 &  0.11\% & 68.4\%   \\
    4 &  1.30\% & 68.4\%   \\
    6 &  1.78\% & 68.4\%   \\
    8 &  1.59\% & 68.4\%   \\
    \bottomrule
  \end{tabular}
  \label{tab:tableResP1S}
\end{table}

Furthermore, the results obtained using a $S_8$ discretization for the angular quadrature and changing the number of cells in the mesh is shown in \ref{tab:tableResP1X}. It is observed that the error augments as the number of meshes augments due to the error build up phenomena previously addressed. In this sense, the less cells the mesh have, the less times information is upstreamed through neural networks, and, hence, the less build up error is produced. It is observed that the influence of the build-up in the error approximately stabilizes for $N \geq 6$. This is because for smaller cells, the dissipation induced by the random errors in the weights of the ANN becomes more important. Moreover, the mean absolute errors obtained for this case are less than 2\% for all cases, which is completely acceptable regarding the current requirements.

\begin{table}
 \caption{Error and acceleration generated by replacing Gauss Elimination with an ANN in Problem 1 for an angular quadrature of order $S_8$}
  \centering
  \begin{tabular}{lllll}
    \toprule
    \multicolumn{5}{c}{Results Problem 1}  \\
    \cmidrule(r){1-3}
    \cmidrule(r){4-5}
    $N_x$ & $N_y$ & $N_z$ & Error ANN & Acceleration  \\
    \midrule
    10 & 10 & 10 & 0.46\% & 72.3\%   \\
    25 & 25 & 25 & 1.05\% & 71.8\%   \\
    50 & 50 & 50 & 1.48\% & 70.6\%   \\
    100 & 100 & 100 & 1.59\% & 68.4\%   \\
    \bottomrule
  \end{tabular}
  \label{tab:tableResP1X}
\end{table}

\subsection{Problem 2: Incident flux in a light absorbing medium}

\begin{figure}
  \centering
  \includegraphics[scale=0.4]{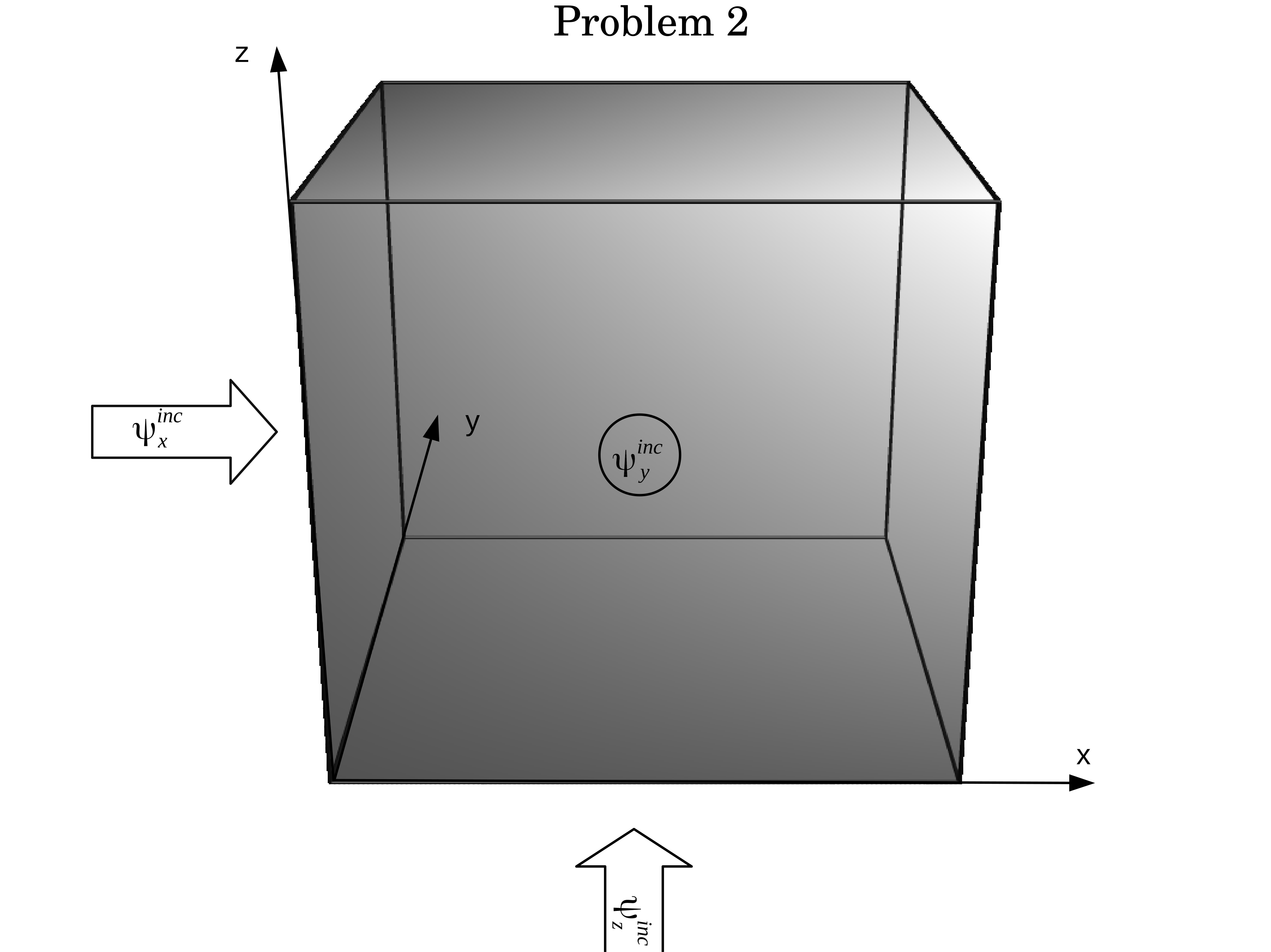}
  \caption{Problem 2.}
  \label{fig:fig2}
\end{figure}

As previously mentioned, Problem 2 consists of an incident flux on a light absorbing medium. The parameters for Problem 2 are shown in Table~\ref{tab:tableP2}. A sketch of the problem is shown in Figure~\ref{fig:fig2}. Incident fluxes of $(\psi_x^{inc}, \psi_y^{inc}, \psi_z^{inc})$ are imposed in the left, back, and bottom faces respectively. Void boundary conditions are imposed in the opposite right, front, and top faces. There are no internal sources in the domain. The total size of the domain for this problem is between $[0.01, 0.017]$ mean free paths. Hence, it is expected that the incident fluxes will be only slightly attenuated while streaming through the domain. Since there is no scattering in this problem, no source iterations are performed and the problem converges one all cells have been swept for all directions. The error function to compare Gauss Elimination with the solutions obtained with the ANN is defined in a similar way than the Problem 1 \ref{Test:Eq2}.

The first test, once again, consists of changing the discretization in angle of the problem, while keeping the spatial discretization fixes with $(\Delta x, \Delta y, \Delta z) = (0.1, 0.1, 0.1)$cm. The results obtained for an angular discretization using using $S_2$, $S_4$, $S_6$ and $S_8$ are presented in Table \ref{tab:tableResP2S}. It is observed that for $S_2$ the mean absolute error obtained are smaller than the one in the previous case. This is because, the relative influence of the error introduced by the neural network presents less influence in this because the angular fluxes are only mildly attenuated.

It can be further observed in Table~\ref{tab:tableResP2S} that the performance of the ANN is deteriorated as the number of sweeping direction per octant are increased. Once again, this is logical since the variance in the input vector of the ANN is increased. However, as was observed in the previous case, it is noted that this phenomenon approximately stabilizes for $N \geq 6$. It is observed that the performance in this case is more rapidly deteriorated than the one in Problem 1 when increasing the number of directions per octant. This is logical since this case presents a larger amount of streaming and, hence, the expected variance in the output of the ANN will be more sensitive to the variance in its input. Nonetheless, the mean errors obtained in these conditions are of only slight larger than 2\%, which is also acceptable for the present requirements.

Furthermore, the results obtained using a $S_8$ discretization for the angular quadrature and changing the number of cells in the mesh is shown in \ref{tab:tableResP2X}. It is observed that the error increases as the number of meshes augments. In this case, this is mainly due to errors produced when evaluating the fluxes in the streaming process. The errors in evaluating the upwind fluxes will propagate to downstream cells, producing a similar phenomenon than the error build up previously observed. However, it is observed that this build up stabilized for large discretization in the angle and the errors obtained are, once again, within the acceptable limits.

\begin{table}
 \caption{Parameters of Problem 2}
  \centering
  \begin{tabular}{llllllllllll}
    \toprule
    \multicolumn{3}{c}{Incident Fluxes $[n/{cm^2s}]$} & \multicolumn{3}{c}{Medium Properties} & \multicolumn{3}{c}{Lengths $[cm]$}  & \multicolumn{3}{c}{Discretization}  \\
    \cmidrule(r){1-3}
    \cmidrule(r){4-6}
    \cmidrule(r){7-9}
    \cmidrule(r){10-12}
    $\psi^{inc}_x$ & $\psi^{inc}_y$ & $\psi^{inc}_z$ &  $\Sigma_{TOT} [1/cm]$ & $\Sigma_S [1/cm]$ & $q [n/(cm^3s)]$ & $L_x$ & $L_y$ & $L_z$ & $N_x$ & $N_y$ & $N_z$  \\
    \midrule
    1.0 &  1.0 &  1.0 &  0.001 &  0 &  0 &  10.0 &  10.0 &  10.0 &  100 &  100 &  100 \\
    \bottomrule
  \end{tabular}
  \label{tab:tableP2}
\end{table}

\begin{table}
 \caption{Error and acceleration generated by replacing Gauss Elimination with an ANN in Problem 1 for a mesh of $100 \time 100 \times 100$ cells}
  \centering
  \begin{tabular}{lll}
    \toprule
    \multicolumn{3}{c}{Results Problem 2}  \\
    \cmidrule(r){1-3}
    Order $S_N$ & Error ANN & Acceleration  \\
    \midrule
    2 &  0.07\% & 68.4\%   \\
    4 &  1.09\% & 68.4\%   \\
    6 &  2.57\% & 68.4\%   \\
    8 &  2.45\% & 68.4\%   \\
    \bottomrule
  \end{tabular}
  \label{tab:tableResP2S}
\end{table}

\begin{table}
 \caption{Error and acceleration generated by replacing Gauss Elimination with an ANN in Problem 2 for an angular quadrature of order $S_8$}
  \centering
  \begin{tabular}{lllll}
    \toprule
    \multicolumn{5}{c}{Results Problem 2}  \\
    \cmidrule(r){1-3}
    \cmidrule(r){4-5}
    $N_x$ & $N_y$ & $N_z$ & Error ANN & Acceleration  \\
    \midrule
    10 & 10 & 10 & 1.23\% & 72.4\%   \\
    25 & 25 & 25 & 1.85\% & 71.8\%   \\
    50 & 50 & 50 & 2.12\% & 70.5\%   \\
    100 & 100 & 100 & 2.45\% & 68.4\%   \\
    \bottomrule
  \end{tabular}
  \label{tab:tableResP2X}
\end{table}

\subsection{Problem 3: Incident flux in an absorbing medium with an internal source}

\begin{figure}
  \centering
  \includegraphics[scale=0.4]{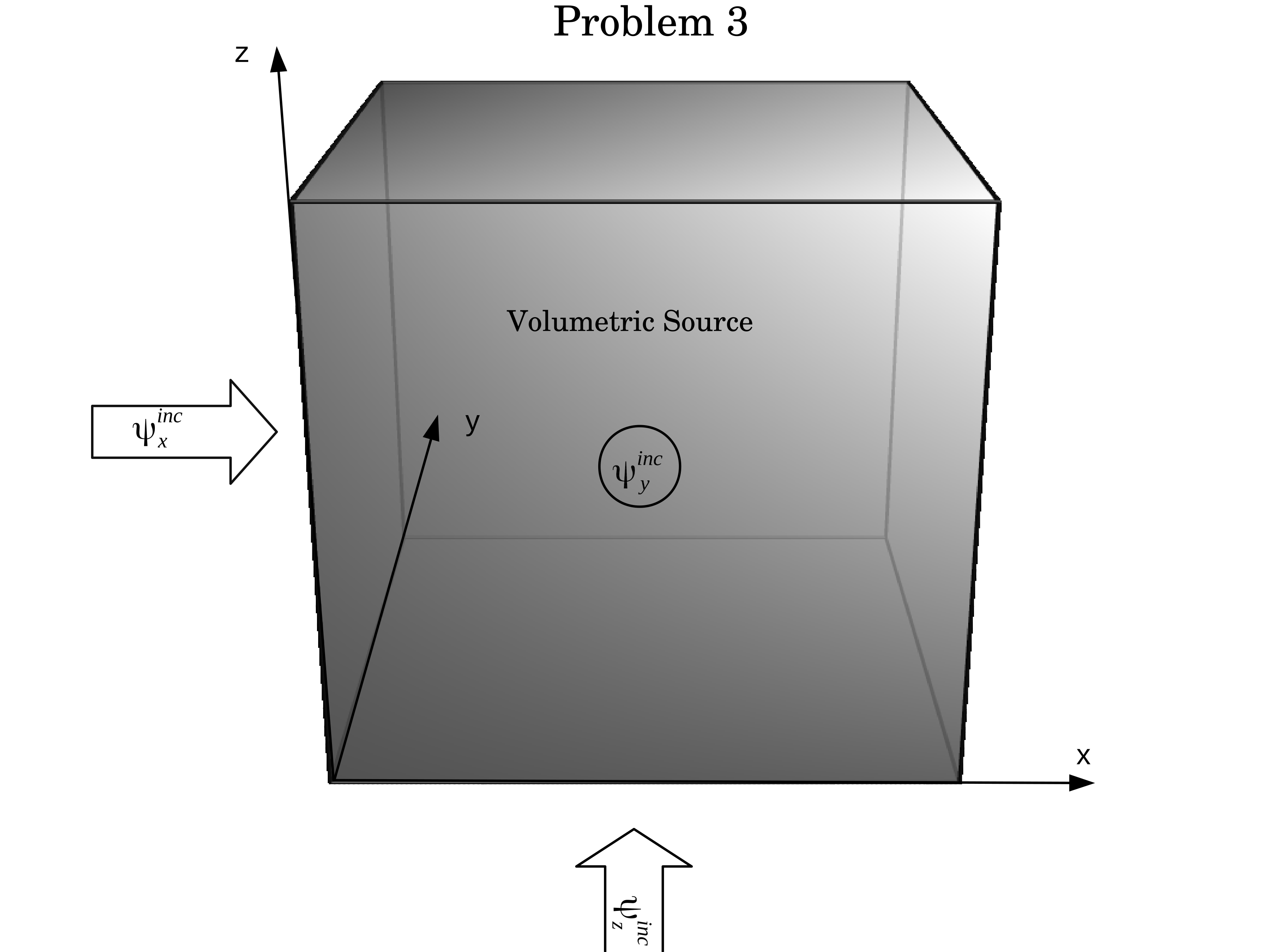}
  \caption{Problem 3.}
  \label{fig:fig3}
\end{figure}

As previously mentioned, Problem 3 consists of an incident flux on a strong absorbing medium. The parameters for Problem 3 are shown in Table~\ref{tab:tableP3}. A sketch of the problem is shown in Figure~\ref{fig:fig3}. Incident fluxes of $(\psi_x^{inc}, \psi_y^{inc}, \psi_z^{inc})$ are imposed in the left, back, and bottom faces respectively. Void boundary conditions are imposed in the opposite right, front, and top faces. There is a constant volumetric internal source in the domain. The total size of the domain for this problem is between $[100, 173]$ mean free paths. Hence, it is expected that the incident fluxes will be strongly attenuated as they travel through the domain. Moreover, it is expected that the emissions of the sources will be bounded over a domain close to the volume of emission. Since there is no scattering in this problem, no source iterations are performed and the problem converges one all cells have been swept for all directions. The error function to compare Gauss Elimination with the solutions obtained with the ANN is defined in a similar way than the Problem 1 \ref{Test:Eq2}.

The first test, once again, consists of changing the discretization in angle of the problem, while keeping the spatial discretization fixes with $(\Delta x, \Delta y, \Delta z) = (0.1, 0.1, 0.1)$cm. The results obtained for an angular discretization using using $S_2$, $S_4$, $S_6$ and $S_8$ are presented in Table \ref{tab:tableResP3S}. It is observed that the errors obtained are in general smaller than the ones obtained in Problem 1. Note that this does not mean that the small error build up effect disappeared but this small error is now reduced in the definition of the error function \ref{Test:Eq2} by a larger intensity of the scalar flux due to the presence of the source. Furthermore, the results obtained using a $S_8$ discretization for the angular quadrature and changing the number of cells in the mesh is shown in \ref{tab:tableResP3X}. Once again it is observed that the mean absolute errors are reduced due to the presence of a larger scalar flux produced by the source. Note finally, that the results obtained for this test problems are within the expected limits of accuracy.

\begin{table}
 \caption{Parameters of Problem 3}
  \centering
  \begin{tabular}{llllllllllll}
    \toprule
    \multicolumn{3}{c}{Incident Fluxes $[n/{cm^2s}]$} & \multicolumn{3}{c}{Medium Properties} & \multicolumn{3}{c}{Lengths $[cm]$}  & \multicolumn{3}{c}{Discretization}  \\
    \cmidrule(r){1-3}
    \cmidrule(r){4-6}
    \cmidrule(r){7-9}
    \cmidrule(r){10-12}
    $\psi^{inc}_x$ & $\psi^{inc}_y$ & $\psi^{inc}_z$ &  $\Sigma_{TOT} [1/cm]$ & $\Sigma_S [1/cm]$ & $q [n/(cm^3s)]$ & $L_x$ & $L_y$ & $L_z$ & $N_x$ & $N_y$ & $N_z$  \\
    \midrule
    1.0 &  1.0 &  1.0 &  10.0 &  0 &  1.0 &  10.0 &  10.0 &  10.0 &  100 &  100 &  100 \\
    \bottomrule
  \end{tabular}
  \label{tab:tableP3}
\end{table}

\begin{table}
 \caption{Error and acceleration generated by replacing Gauss Elimination with an ANN in Problem 3 for a mesh of $100 \time 100 \times 100$ cells}
  \centering
  \begin{tabular}{lll}
    \toprule
    \multicolumn{3}{c}{Results Problem 3}  \\
    \cmidrule(r){1-3}
    Order $S_N$ & Error ANN & Acceleration  \\
    \midrule
    2 &  0.14\% & 68.4\%   \\
    4 &  0.86\% & 68.4\%   \\
    6 &  0.68\% & 68.4\%   \\
    8 &  0.45\% & 68.4\%   \\
    \bottomrule
  \end{tabular}
  \label{tab:tableResP3S}
\end{table}

\begin{table}
 \caption{Error and acceleration generated by replacing Gauss Elimination with an ANN in Problem 3 for an angular quadrature of order $S_8$}
  \centering
  \begin{tabular}{lllll}
    \toprule
    \multicolumn{5}{c}{Results Problem 3}  \\
    \cmidrule(r){1-3}
    \cmidrule(r){4-5}
    $N_x$ & $N_y$ & $N_z$ & Error ANN & Acceleration  \\
    \midrule
    10 & 10 & 10 & 0.27\% & 72.4\%   \\
    25 & 25 & 25 & 0.38\% & 71.8\%   \\
    50 & 50 & 50 & 0.52\% & 70.5\%   \\
    100 & 100 & 100 & 0.45\% & 68.4\%   \\
    \bottomrule
  \end{tabular}
  \label{tab:tableResP3X}
\end{table}

\subsection{Problem 4: Incident flux in an absorbing and scattering medium with an internal source}

\begin{figure}
  \centering
  \includegraphics[scale=0.4]{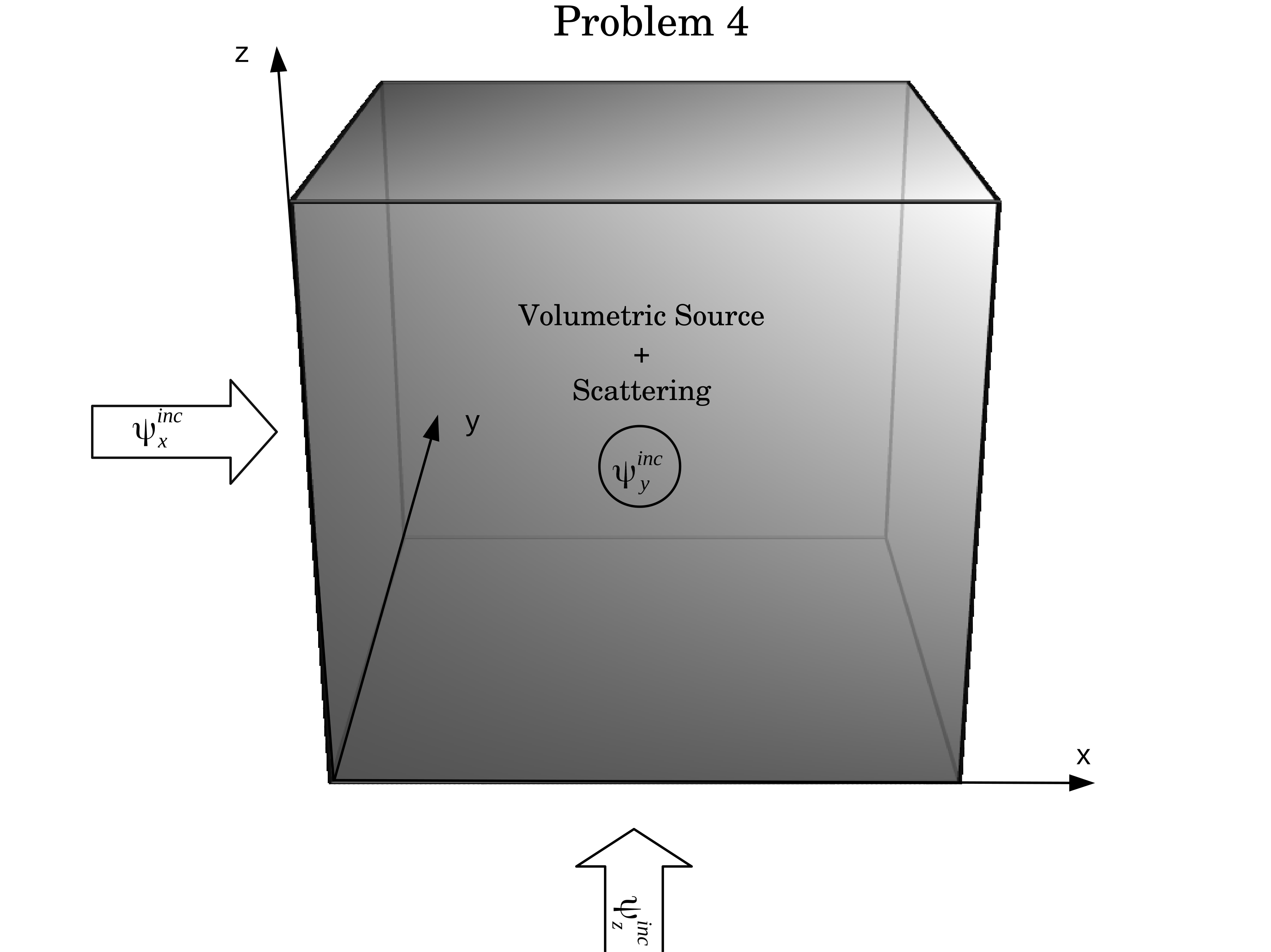}
  \caption{Problem 4.}
  \label{fig:fig4}
\end{figure}

As previously mentioned, Problem 4 consists of an incident flux on a strong absorbing medium with s large scattering ratio. The parameters for Problem 4 are shown in Table~\ref{tab:tableP4}. A sketch of the problem is shown in Figure~\ref{fig:fig4}. Incident fluxes of $(\psi_x^{inc}, \psi_y^{inc}, \psi_z^{inc})$ are imposed in the left, back, and bottom faces respectively. Void boundary conditions are imposed in the opposite right, front, and top faces. There is a constant volumetric source in the domain. The total size of the domain for this problem is between $[100, 170]$ mean free paths. Note, however, that even though the mean free path of radiation is much smaller than the domain, most of the interactions produced are by scattering. Hence, a build up of the flux within the domain is expected. Since this problem has a high-scattering ratio several source iterations will be necessary until convergence. The error function to compare Gauss Elimination with the solutions obtained with the ANN is defined in a similar way than the Problem 1 \ref{Test:Eq2}. For convergence criterion on four source iterations as follows:

\begin{equation}
    \epsilon_S = \frac{1}{N_x N_y N_z} \sum_{i=1}^{N_x} \sum_{j=1}^{N_y} \sum_{k=1}^{N_z} \frac{|\phi^{(n)}(x_i,x_j,x_k) - \phi^{(n-1)}(x_i,x_j,x_k)|}{|\phi^{(n)}(x_i,x_j,x_k)|},
    \label{Test:Eq3}
\end{equation}

where $\phi^{(n)}(x_i,x_j,x_k)$ is the scalar flux computed in the current source iteration either by Gauss Elimination or the ANN and $\phi^{(n-1)}(x_i,x_j,x_k)$ is the one computed in the previous iteration. Source iterations were finished when $\epsilon_s < 10^{-4}$.

\begin{table}
 \caption{Parameters of Problem 4}
  \centering
  \begin{tabular}{llllllllllll}
    \toprule
    \multicolumn{3}{c}{Incident Fluxes $[n/{cm^2s}]$} & \multicolumn{3}{c}{Medium Properties} & \multicolumn{3}{c}{Lengths $[cm]$}  & \multicolumn{3}{c}{Discretization}  \\
    \cmidrule(r){1-3}
    \cmidrule(r){4-6}
    \cmidrule(r){7-9}
    \cmidrule(r){10-12}
    $\psi^{inc}_x$ & $\psi^{inc}_y$ & $\psi^{inc}_z$ &  $\Sigma_{TOT} [1/cm]$ & $\Sigma_S [1/cm]$ & $q [n/(cm^3s)]$ & $L_x$ & $L_y$ & $L_z$ & $N_x$ & $N_y$ & $N_z$  \\
    \midrule
    1.0 &  1.0 &  1.0 &  10.0 &  9.9 &  1.0 &  10.0 &  10.0 &  10.0 &  100 &  100 &  100 \\
    \bottomrule
  \end{tabular}
  \label{tab:tableP4}
\end{table}

The first test, once again, consists of changing the discretization in angle of the problem, while keeping the spatial discretization fixes with $(\Delta x, \Delta y, \Delta z) = (0.1, 0.1, 0.1)$cm. The results obtained for an angular discretization using using $S_2$, $S_4$, $S_6$ and $S_8$ are presented in Table \ref{tab:tableResP4S}. It is observed that errors comparable to Problem 3 are obtained. Once again, this is because there is a larger flux in the domain, which decreases the relevance of the build up error in \ref{Test:Eq1}. A similar behaviour is observed in \ref{tab:tableResP4X} for the discretization in space, where the errors are comparable to the ones obtained in Problem 3. All the errors obtained in this case are within the acceptable limits.

One further interesting phenomenon observed is that the solutions obtained with ANNs converge in less source iterations than the ones obtained with Gauss Elimination. In fact, the effective spectral radius of system \ref{Eq:T12} is approximately ~0.957 when computing the solution with Gauss Elimination but is of ~0.859 when computing the solutions with the ANNs. This behaviour was explained since the noise present in the connections of the network diffuses through the domain reducing directional the marked directionality of the $S_N$ transport process and, hence, accelerating the convergence in source iterations. To measure this acceleration, the overall acceleration has been defined as the time to compute the solution by Gauss Elimination divided by the time to compute the solution with ANNs in Tables \ref{tab:tableResP4S} and \ref{tab:tableResP4X}. It is observed that an overall acceleration factor of ~12 is obtained in the present case. However, note that this acceleration factor will be reduced as the scattering ratio is reduced, going asymptotically to the speed-up introduced by the ANN in the solution process as the scattering ratio goues to zero.

\begin{table}
 \caption{Error and acceleration generated by replacing Gauss Elimination with an ANN in Problem 4}
  \centering
  \begin{tabular}{llllll}
    \toprule
    \multicolumn{6}{c}{Results Problem 4}  \\
    \cmidrule(r){1-6}
    Order $S_N$ & Error ANN & \makecell{DGFEM \\ Source Iterations} & \makecell{ANN \\ Source Iterations} & \makecell{Iteration \\ Acceleration} & \makecell{Overall  \\ Acceleration}  \\
    \midrule
    2 &  0.44\% & 211 & 101 & 81.9\% & 11.4 \\
    4 &  0.39\% & 220  & 102 & 81.7\%  & 11.8\\
    6 &  0.87\% & 228 & 104 & 81.4\% & 12.0\\
    8 &  0.87\% & 237  & 110 & 81.7\% & 11.7\\
    \bottomrule
  \end{tabular}
  \label{tab:tableResP4S}
\end{table}

\begin{table}
 \caption{Error and acceleration generated by replacing Gauss Elimination with an ANN in Problem 4 for an angular quadrature of order $S_8$}
  \centering
  \begin{tabular}{llllllll}
    \toprule
    \multicolumn{8}{c}{Results Problem 4}  \\
    \cmidrule(r){1-3}
    \cmidrule(r){4-8}
    $N_x$ & $N_y$ & $N_z$ & Error ANN & \makecell{DGFEM \\ Source Iterations} & \makecell{ANN \\ Source Iterations} & \makecell{Iteration \\ Acceleration} & \makecell{Overall  \\ Acceleration}  \\
    \midrule
    10 & 10 & 10 & 0.56\% & 15  & 9  & 83.1\% & 9.1   \\
    25 & 25 & 25 & 0.73\% & 42  & 20  & 82.9\% & 11.4   \\
    50 & 50 & 50 & 0.94\% & 121  & 58 & 82.4\% & 11.4  \\
    100 & 100 & 100 & 0.87\% & 237 & 110 & 81.7\% & 11.7  \\
    \bottomrule
  \end{tabular}
  \label{tab:tableResP4X}
\end{table}




\section{Conclusions}

The present works deals with the development of a Machine Learning technique for accelerating radiation transport. In the present work, radiation transport is solved numerically with a $S_n$ discretization in angle and discontinuous finite elements in space. This discretization yields a set of small matrix-vector systems that have to be solved for the angular radiation fluxes for cell in the spatial domain and for each sweeping direction in the angular domain. Hence, the problem of finding the angular fluxes is localized to each cell and direction. The key idea motivating this work was that the systems obtained for these one-cell one-direction problems have a certain structure which is learned by a Machine Learning approach. Then, the resulting artificial neural network is employed  to solve more rapidly the local linear systems, generated for each cell in the domain and direction in the angular quadrature.

For this purpose, the discretization process of the transport equation have been first analyzed in detail in order to show how the structure in these matrix-vector systems appears. This analysis motivated the usage of Artificial Neural Networks (ANNs) as a Machine Learning approach to map the inputs of the one-cell problems into the angular fluxes over this sell, i.e. to solve the one cell problems. The design process for the ANNs was subjected to two competing constraints. First, the regression error in the ANNs should be small for the one-cell transport solutions used to train these ANNs. Then, the number of connections in the ANN should be reduced, to improve their evaluation speed and, hence, improve the speed of transport solvers when implementing the ANNs as their solution kernels. Thus, special attention was given to the design of these ANNs. 

These ANNs have only one hidden layer in order to reduce the number of operations required for their feed-forward evaluation. Moreover, scaling of the input values and the activation function in the hidden layer have been optimized for linearizing output of this hidden layer, while preserving the linear behaviour of the transport problem. Moreover, the output activation function have been optimized to include full support over positive real numbers and to parametrically filter large negative solutions. In addition, a Dropout technique have been tailored to reduce the number of connections in the network for improving its feed-forward evaluation efficiency. Finally, a sensitivity study was performed to verify that the regression response of the developed network was similar to the one obtained with ANNs with more hidden layers and hidden neurons. The networks obtained have been trained on a set of random one-cell transport problems with average training and testing accuracy of ~$10^{-4}$. Moreover, the final architecture of ANN proposed allowed to reduce the number of computational operations required for its feed-forward evaluation by a factor of 5 compared to the solution by Gauss Elimination.

The ANNs were then included into a 3D transport solver and tested against the solutions provided by regular Gauss Elimination solver. Four different 3D test problems were selected for this purpose, which allowed to incrementally test the performance of the ANN. It was observed that a buildup in the error was obtained when sweeping the solution process through the domain. However, in transport problems with large fluxes, the built up error is not large compared to the values obtained for the fluxes (~$2\%$). Moreover, a regularizer was introduced in the cost function of the ANN in order to limit the influence of this error when the fluxes are small and, hence, errors of less than $2\%$ were obtained for thick absorbers.

There are several perspectives for the present work. For instance, the solutions provided by ANNs can be tested on neutronic benchmark problems in order to assess their performance in non-homogeneous domains. Moreover, the activation functions in the ANN can be computationally optimized in order to improve the regression performance of ANNs. In addition, the train and test sets could be replace with exact solutions to one cell problems, allowing the ANN to learn form more accurate solutions. Furthermore, ANNs can be used as low order operators in order to reduce their influence over the global error. It is worth to mention that this are just some few of the plethora of possibilities that ANNs presents for improving the efficiency and accuracy in radiaiton transport.

\bibliographystyle{unsrt}

\end{document}